\documentclass[letterpaper, 10 pt, conference]{ieeeconf}

\IEEEoverridecommandlockouts                              

\usepackage{amsmath,amsfonts}
\usepackage{algpseudocode}
\usepackage{algorithm}
\usepackage{array}
\usepackage[caption=false]{subfig}
\usepackage{tabularx}
\usepackage{colortbl}
\usepackage{textcomp}
\usepackage{stfloats}
\usepackage[normalem]{ulem}
\usepackage{multirow}
\usepackage{verbatim}
\usepackage{graphicx}
\usepackage{cite}
\usepackage{flushend}
\usepackage{diagbox}
\usepackage[dvipsnames]{xcolor}
\usepackage{threeparttable}
\usepackage{booktabs}
\usepackage{empheq}
\usepackage{makecell}
\usepackage{url}
\usepackage{soul}
\usepackage{tikz}
\usepackage{gensymb}
\definecolor{redvw}{HTML}{F36B9C}
\definecolor{bluevw}{HTML}{1E88E5}
\definecolor{greenvw}{HTML}{005648}

\usepackage{setspace}
\usepackage[colorlinks=true, linkcolor=black, citecolor=black, urlcolor=blue]{hyperref}
\usepackage{cleveref}
\crefname{figure}{Fig.}{Figs.}
\Crefname{figure}{Figure}{Figures}

\begin{document}

\title{\LARGE \bf PalpAid: Multimodal Pneumatic Tactile Sensor for Tissue Palpation
}

\author{Devi Yuliarti \textsuperscript{1*},
Ravi Prakash \textsuperscript{1*$^{\dagger}$},
Hiu Ching Cheung \textsuperscript{2},
Amy Strong \textsuperscript{1},
Patrick J. Codd \textsuperscript{1}, 
Shan Lin \textsuperscript{2}
 \thanks{\href{https://raprakashvi.github.io/palpaid/}{Project Website}}
 \thanks{\textsuperscript{*}Equal contribution. \textsuperscript{1}Duke University. \textsuperscript{2}Arizona State University.}%
\thanks{This work was partially supported by the Duke Bass Connections Award.}%
\thanks{\textsuperscript{$\dagger$}Corresponding author: Ravi Prakash (\texttt{ravi.prakash@duke.edu}).}%
}

\maketitle
\begin{abstract}
The tactile properties of tissue, such as elasticity and stiffness, often play an important role in surgical oncology when identifying tumors and pathological tissue boundaries. Though extremely valuable, robot-assisted surgery comes at the cost of reduced sensory information to the surgeon, with vision being the primary. Sensors proposed to overcome this sensory desert are often bulky, complex, and incompatible with the surgical workflow. We present PalpAid, a multimodal pneumatic tactile sensor to restore touch in robot-assisted surgery. PalpAid is equipped with a microphone and pressure sensor, converting contact force into an internal pressure differential. The pressure sensor acts as an event detector, while the acoustic signature assists in tissue identification. We show the design, fabrication, and assembly of sensory units with characterization tests for robustness to use, repetition cycles, and integration with a robotic system. Finally, we demonstrate the sensor's ability to classify 3D-printed hard objects with varying infills and soft \textit{ex vivo} tissues. We envision PalpAid to be easily retrofitted with existing surgical/general robotic systems, allowing soft tissue palpation. 

\end{abstract}

\section{Introduction}
The sense of touch is fundamental in our ability to perceive and identify objects. Clinicians regularly employ palpation as feedback for initial assessment for cancer diagnostics \cite{gorman2021importance,shetty2015accuracy} based on the contour, composition, and stiffness of tissues. %
%
The advent of robot-assisted minimally invasive surgery (RAMIS) has been transformative for improvements in postoperative patient care. It offers dexterous and precise operation through minimal incision, leading to reduced complication rates and enhanced patient outcomes. However, the minimally invasive nature prevents surgeons from directly palpating tissues and limits the surgeon's access to crucial sensory feedback. While popular platforms like da Vinci 5 \cite{moschovas2024first} have recently integrated force feedback, it is limited to a coarse resolution that could be insufficient for subtle changes in tissue composition.

Works using vision-based force estimation exist, where a mapping from 3D tissue deformation to applied force is learned \cite{su2019multicamera}. However, prior knowledge of tissue parameters is typically needed, which is often unavailable. Developments in vision-based tactile sensors (VBTS) have enabled sub-millimeter spatial resolution and have combined modalities such as force, texture, and temperature for providing rich contact information \cite{yuan2017gelsight, zhang2025utact,tiong2025omnisense,prince2025tacscope}. 
While helpful, the dependence on vision and the general need for illumination render the sensor complex and bulky. 
Works \cite{prince2025tacscope, di2024using,kapuria2024robot} have proposed VBTS with potential for RAMIS applications, leveraging optically compatible design with computer vision algorithms to achieve tumor classification and shape estimation while keeping the form factor small. Yet, the presence of electronic components and multilayer design limits clinical adoption due to large form factor, high cost, and incompatibility with sterilization. 

Several recent works have leveraged acoustics as a modality to differentiate materials under vision occlusion or absence \cite{liu2025sonicsense, chen2022boombox}, as materials vibrate at their characteristic natural frequency, which is often a strong indicator of material type \cite{wall2023passive,li2025acoustac,liu2025wildfusion}. Wall et al., \cite{wall2023passive} showed the potential of using off-the-shelf MEMS microphones in confined pneumatic cavities to detect external contact location, force, and temperature. Since then, various works have used acoustic waves for force estimation \cite{mandil2024acoustic}, to support vision \cite{andrussow2025adding}, in optoacoustic sensing \cite{bao2025miniaturized}, and with a force sensor for material boundary (fat vs muscle) identification \cite{chen2025boundary}.

\begin{figure}[t] 
    \centering
    \vspace{-2mm} 
    \includegraphics[width=\linewidth]{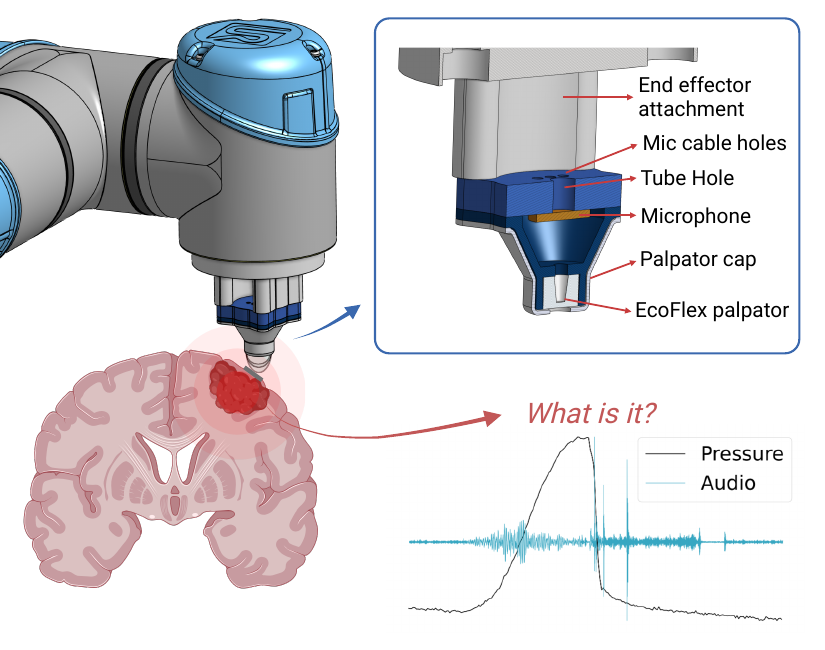}
    \vspace{-2mm} 
    \caption{\textbf{PalpAid:} Multimodal pneumatic tactile sensor with adaptive palpator for compliance with soft tissue. The presence of a microphone and a pressure sensor in a connected confined air cavity allows for extracting rich tool-tissue contact interaction, allowing tissue delineation in a surgical setting \cite{prakash2025biorender}.}
    \label{fig:intro_figure}
    \vspace{-10pt}
\end{figure}



We present PalpAid (\cref{fig:intro_figure}), a multimodal pneumatic tactile sensor that leverages the internal air pressure differential created through tool-tissue interaction for tissue identification. The sensor consists of an enclosed microphone inside a sealed air cavity with an expandable layer of silicone. A pressure sensor is attached in line with the air channel. In an inflated state, contact-induced force variation creates internal pressure changes that are picked up by the microphone with air as the conduction medium. The pressure sensor measurement acts as an indicator for contact-based event detection, allowing filtering of temporally synced acoustic data for the duration of contact. Overall, our contributions can be summarized as:
\begin{itemize}[left=0pt]
    \item \textbf{Modular \& Multimodal}: Design and development of a modular, multimodal pneumatic tactile sensor converting contact force to an internal pressure differential.
    \item \textbf{Material Classification}: We validate the effectiveness of PalpAid on hard 3D-printed material as well as \textit{ex vivo} soft-tissue specimens from various animals.
    \item \textbf{Multipurpose}: The palpator adapts its dimensions as a function of internal pressure -- allowing compliance with tissues of varied stiffness. 
    \item \textbf{Clinically compatible}: Minimal components, small form factor, and low-cost manufacturing allow easy sterilization through simply replacing the palpator head.

\end{itemize}


\begin{figure*}[!t] 
    \centering
    \vspace{-2mm} 
    \includegraphics[width=\textwidth]{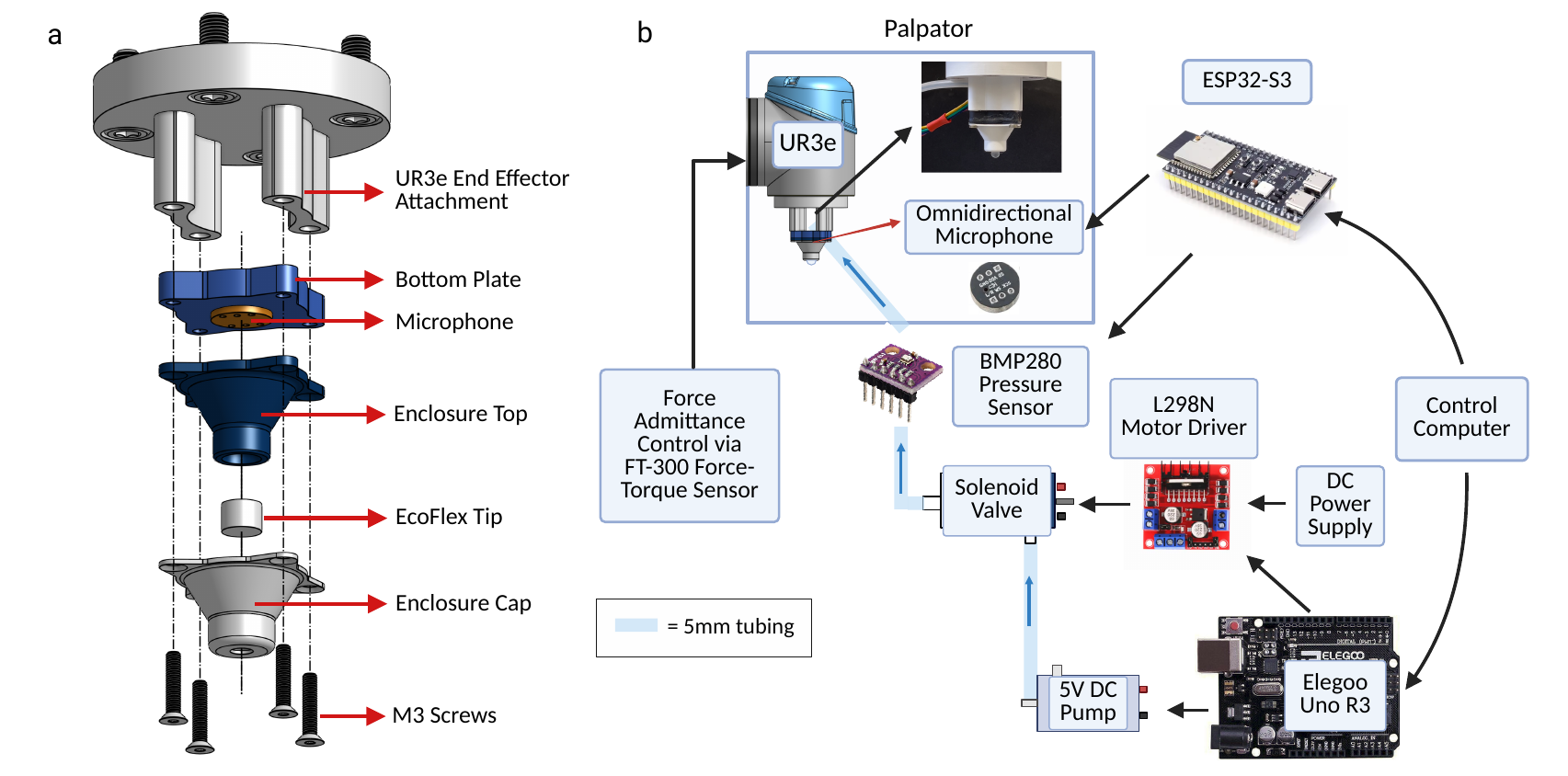}
    \vspace{-2mm} 
    \caption{\textbf{System Overview}: (a) Exploded view of PalpAid assembly with major components highlighted, (b) System architecture diagram of robot-mounted sensor with data acquisition and pressure control components.} 
    \label{fig:system_overview}
    \vspace{-10pt}
\end{figure*}

 Unlike previous work that employed a microphone in confined air cavities under active and passive sensing \cite{wall2023passive}, PalpAid uses passive acoustic waves generated by the deformation of the palpator in a very small air cavity. Because of its modular nature, this design has potential for miniaturization and future adaptation to existing commercial robots and tools, including surgical trocars. We envision PalpAid to use the rich and complementary features captured by a sensitive MEMS microphone and pressure sensors to distinguish between different tissue and tumor types while being gentle to the natural shape of the soft tissue. 
 %
 




\section{Methods}

PalpAid is designed to leverage acoustic and tactile sensing when interacting with materials. We aim to assist the surgical workflow with a compliant, yet effective, sensory data stream to detect and delineate tissues of varied stiffness via contact.

\subsection{Device Overview}

The overall assembly and architecture of the PalpAid sensor is shown in \Cref{fig:system_overview}. The sensor is attached to a UR3e robotic manipulator. The sensor contains a silicone tip, known as the palpator, that interacts with the sample. The tip is inflated to maintain an internal pressure of choice. The internal pressure, and, subsequently, the palpator dimension can be changed to adjust for tissue stiffness. When the palpator makes contact with an object, the silicone deforms, changing the internal pressure and volume of the palpator cavity to produce a sound.
 An omnidirectional microphone contained within the palpator enclosure records the vibrations produced by this contact.

The pneumatic system consists of a 5V DC motor pump, 6V solenoid valve, BMP280 pressure sensor (Bosch Sensortec, Germany), and a pressure gauge. The pump and solenoid valve are controlled by the L298N motor driver. A high signal-to-noise (SNR) mono microphone, ICS43434 (TDK Corporation, Japan), is used. Data from the pressure sensor and microphone are collected using an ESP32-S3 N16R8 microcontroller (Espressif Systems, China). The sensor is attached to a UR3e robotic manipulator with an FT 300-S Force Torque Sensor (Robotiq, Canada) at its end effector to ensure consistent contact force during sensing. 


{\color{blue}
\begin{table}[t]
\caption{Prototype Parameters}
\centering
\scriptsize
\setlength{\tabcolsep}{3pt}
\renewcommand{\arraystretch}{0.9}
\begin{tabular}{p{0.30\columnwidth} p{0.45\columnwidth} r}
\toprule
\textbf{Component} & \textbf{Parameter} & \textbf{Dimensions} \\
\midrule
\multirow{5}{*}{\textit{Enclosure}} &
Height & 25\,mm \\
& Wall thickness & 1.715\,mm \\
& Base inner diameter & 20\,mm \\
& Base outer width & 29.842\,mm \\
& Internal volume & 2142.378\,mm$^{3}$ \\
\midrule
\multirow{2}{*}{\textit{EcoFlex palpator}} &
Thickness & 6\,mm \\
& Diameter & 8.23\,mm \\
\midrule
\multirow{2}{*}{\textit{Microphone PCB}} &
Thickness & 1.6\,mm \\
& Outer radius & 13.5\,mm \\
\midrule
\multirow{2}{*}{\textit{Enclosure cap}} &
Wall thickness & 0.75\,mm \\
& Top hole diameter & 5\,mm \\
\bottomrule
\end{tabular}
\label{tab:prototype_parameters}
\end{table}
}

\begin{table}[t]
\caption{Prototype Costs}
\centering
\scriptsize
\setlength{\tabcolsep}{3pt}
\renewcommand{\arraystretch}{0.9}
\begin{tabular}{p{0.66\columnwidth} r}
\toprule
\textbf{Component} & \textbf{\$ / device} \\
\midrule
Fast ABS-like resin & 0.31 \\
poxy & 0.60 \\
Sil-Poxy & 0.31 \\
Pressure sensor & 1.47 \\
Solenoid valve & 2.95 \\
ICS43434 microphone & 5.33 \\
DC pump + tubing & 7.95 \\
Ecoflex 00-20 & 0.00004 \\
\midrule
\textbf{Total} & \textbf{19.11004} \\
\bottomrule
\end{tabular}
\label{tab:prototype_costs}
\end{table}

\subsection{Fabrication of the Pneumatic Palpator}\label{sec:fabrication}

\begin{figure}[t] 
    \centering
    \vspace{-2mm} 
    \includegraphics[width=\linewidth]{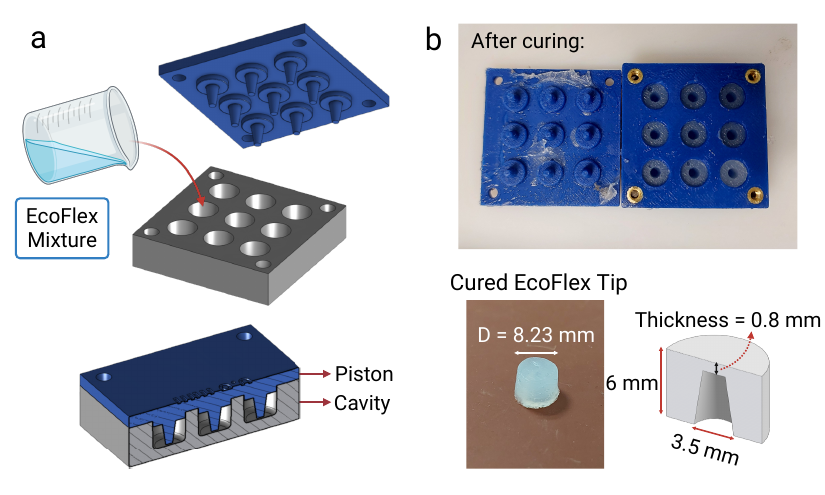}
    \caption{\textbf{Sensor Fabrication}: (a) A 1:1 ratio EcoFlex mixture was poured into a two-part mold, (b) EcoFlex was cured at room temperature for a minimum of 4 hours before integrating it with the palpator enclosure \cite{prakash2025biorender}.}
    \label{fig:fabrication}
    \vspace{-10pt}
\end{figure}

Every component of PalpAid enclosure (\cref{fig:system_overview}a) was Stereolithography (SLA) printed using Acrylonitrile Butadiene Styrene (ABS) thermoplastic polymer resin.
SLA printing allowed $50 \mu$m precision and achieved both a smooth surface finish and better dimensional accuracy than conventional Fused Deposition Modeling (FDM) printing. %
The palpator was fabricated by casting silicone (EcoFlex 00-20, Smooth-On, PA, United States) in a 3D-printed PLA (polylactic acid) two-part mold. Silicone was chosen as the material due to biocompatibility \cite{janardhana2025comprehensive}, high elasticity (845 \%), and low Shore hardness (00-20) \cite{SmoothOn_Ecoflex_00_20}. These properties enable stable, inflated volume across different pressures and multiple palpation cycles. The two-part palpator mold included a tubular cavity (8.23 mm base diameter) and a piston part to ensure the palpator maintains a 0.8 mm tip thickness, as shown in \cref{fig:fabrication}. Before casting, the mold was cleaned with acetone and coated with Smooth-On mold release spray. To cast the palpator, a 1:1 ratio mixture of EcoFlex 00-20 part A and B was vacuum-degassed to release all air bubbles, poured into the mold, and left to cure at room temperature for  $\sim4$ hours. 

Once the palpator was cured, it was securely joined to the enclosure by applying Sil-Poxy (Silicone Epoxy) glue. Epoxy and Sil-Poxy were used because of their mechanical and adhesive properties. The microphone was then inserted into the bottom plate of the palpator, which was later joined and sealed to the top of the enclosure via epoxy resin. Then, $5$~mm OD tubing was connected to the bottom plate of the palpator enclosure to interface with the pneumatics subsystem. Finally, to achieve a small palpator base diameter of $5$~mm, similar to that of a trocar, an offset "cap" with a similar-sized hole on top was attached to the enclosure with four M3 screws.

\begin{figure}[t] 
    \centering
    \includegraphics[width=\linewidth]{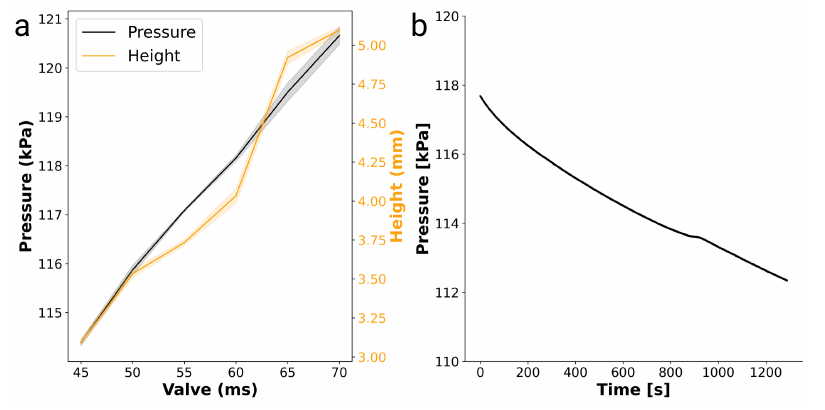}
    \vspace{-2mm} 
    \caption{\textbf{Pressure Stability}: (a) Effect of valve opening time on pressure and resultant palpator height; (b) Gradual pressure reduction with time in relaxed state at palpator initial height of $4$~mm over a span of $20$~minutes.}
    \label{fig:pressure_stability}
    \vspace{-10pt}
\end{figure}

\subsection{Embedded Electronics and Sensing}\label{sec:electronics}
MEMS-based sensors were employed because of their small form factor and low cost. The ESP32-S3 (N16R8) allows for direct audio recording via the $I^2$ (Inter-IC Sound) serial communication protocol. The microphone employed, ICS43434, uses a 24-bit $I^2S$ interface with a 8000 Hz sampling rate for digital data acquisition and has a high SNR and sensitivity. The pressure sensor BMP280 uses $I^2C$ protocol for communicating with ESP32 at a 15.625 Hz sampling rate in sync with the microphone through serial transmission.

A 6V solenoid valve is powered by PWM outputs of the L298N motor driver connected to an Elegoo Uno R3 and a DC power supply, enabling precise control of valve opening time. This is needed to achieve desired pressure values and palpator heights for experimental trials. The input air at a steady flow rate of $2.5$ litres per minute is supplied to the palpator by a miniature 5V DC pump powered by the Elegoo Uno R3.

\subsection{Bench Testing Protocols}\label{sec:calibration_protocols}
Under a zero-load condition, the dimension of the palpator depends on internal air pressure. We assessed the effect of valve opening time on internal pressure and palpator height with height measured from the base of palpator tip.
Pressure values were measured after 10 seconds of inflation. Trials were repeated (n=3) at room temperature and pressure. Valve opening time ($40$~ms - $70$~ms) was chosen experimentally as a reasonable functional range to maintain corresponding palpator height between $3$~mm - $5$~mm (\cref{fig:pressure_stability}a). A palpator height of $\sim4$~mm was chosen as the initial condition for experiments performed downstream. After one full experimental cycle, the palpator (cap) was easily replaced.


\subsection{Stability Over Time}\label{sec:stability_experiment}
To assess the sensor's internal pressure stability over time, pressure data was recorded at a tip height of 4 mm, corresponding to an average initial pressure of 119 kPa, for 20 minutes. 
As seen in \cref{fig:pressure_stability}b, pressure drop of 5.35 kPa (0.0045 kPa/s) was observed over this time period.

\begin{figure*}[!t] 
    \centering
    \vspace{-2mm} 
    \includegraphics[width=\textwidth]{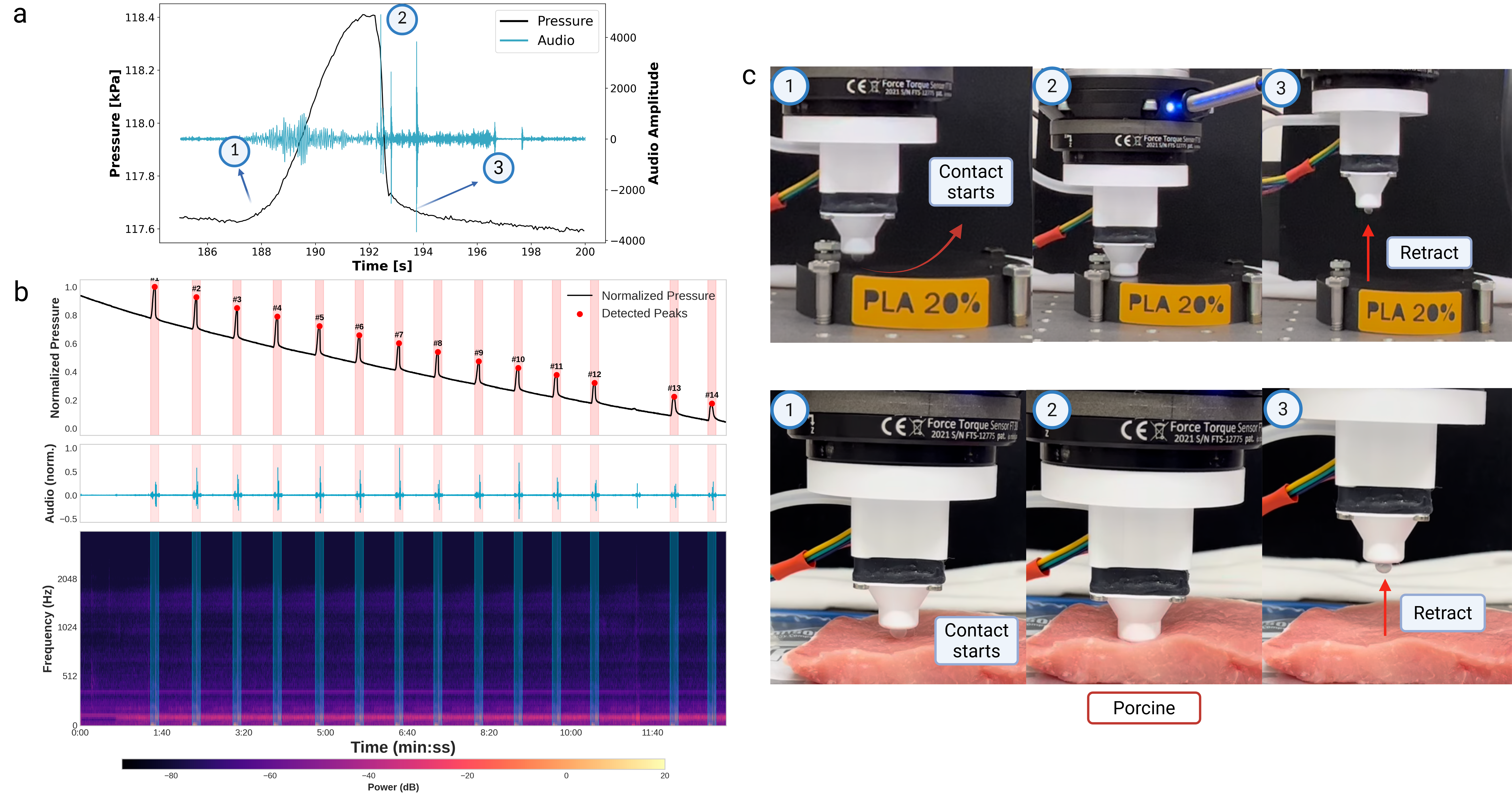}
    \vspace{-2mm} 
    \caption{\textbf{Data}: (a) Single palpation event showing palpator contact with surface (1), full compression with deflation (2), and retraction with inflation (3) ;  (b) Example of time-synced pressure and acoustic data collected over a series of palpation events in pork. The temporal regions extracted for downstream tasks are highlighted based on peaks detected from pressure signal ; (c) Corresponding visuals for PLA $20 \%$ and pork tissue corresponding to (b). }
    \label{fig:Tap_peak}
    \vspace{-15pt}
\end{figure*}

\section{Experiments and Results}\label{sec:experiments}

\subsection{Data Collection}\label{sec:data_collection}

Device performance was tested by evaluating the sensor's ability to classify different materials and tissues from data collected over several cycles: shown in \cref{fig:Tap_peak}a for a single palpation event and \cref{fig:Tap_peak}b for multiple palpation events sequentially. We considered two broad cases: hard-PLA-based and soft animal tissues. For each material type, the aim was to classify the subcategory of material. The hard materials category consisted of 3D printed PLA with infill percentages of 10\%, 15\%, and 20\%. We chose PLA as a flat planar surface to reduce signal differences resulting from the palpator-material attack angle. The \textit{ex vivo} soft animal tissue group consisted of porcine, chicken, and bovine tissue from the supermarket, which represented a variety of biological textures and elasticity (\cref{fig:Tap_peak}c). Consistent with the protocol described in \cref{sec:calibration_protocols}, the sensor was attached to the UR3e robot arm, with $4$~mm palpator height at $118.6$~kPa. Before data collection on a material, the palpator height was reinitialized to $4$~mm. 

During robot-assisted palpation, admittance control was used to control the tapping motion with a preset maximum permissible force and error 
such that the end effector retracts once a set force is reached. Before contact, the tip was inflated and deformed as per contact with the material. Upon reaching target force, the palpator stayed in deflated position for $0.3$~seconds. This method allows for gentle and consistent palpating, but also prevents damage or deformation of the material samples. For softer material, after the palpator deformed completely, the palpator cap was noticed to barely contact the surface, as seen in \cref{fig:Tap_peak}c for porcine. In the future, using pressure differential as feedback over the applied robot end-effector force can mitigate this issue.


\subsection{Data Processing and Training}

During each palpation, the microphone and the pressure sensor connected to the palpator recorded time-synchronized data, shown in \cref{fig:Tap_peak}a. To reduce ambient sensor noise and identify events where contact was detected, pressure sensor data was used to find peaks. Before processing, the pressure values were normalized to remove any artifacts due to leakage. Once the peaks were identified, 10 seconds of audio signal with the detected peak time as the center was extracted.

To represent audio data features, we used the Mel-Frequency Cepstral Coefficients (MFCC). MFCC captures the spectral envelope of audio signals by dividing them into small overlapping time frames to capture time-varying patterns, applying a Mel-scale filter bank, processing the output in a logarithmic scale, and executing the Discrete Cosine Transform (DCT) to get the Cepstral Coefficients \cite{ai2012classification}. This process has been shown to extract relevant high-level acoustic features useful in auditory event classification tasks and serves here as a naive feature set for classification. 

For each palpation cycle, MFCCs (N = 20) were extracted and averaged over the total windows. Each palpation event was assigned a label corresponding to the material. Owing to the small dataset size, a Support Vector Machine (SVM) classifier with a radial basis function kernel was used. For each task, a 5-fold cross-validation test was performed to account for data distribution variability.

\begin{table}[ht]
\centering
\caption{Per-Class Cross-Validation Accuracy Across 5 Folds. 
Samples are shown as $N_{\text{Train}}/N_{\text{Test}}$.}

\label{tab:cv_accuracy}
\renewcommand{\arraystretch}{1.2}
\setlength{\tabcolsep}{10pt}
\begin{tabular}{lcc}
\hline
\textbf{Label} & \textbf{Accuracy (\%)} & \textbf{Samples} \\ 
\hline
PLA $10\%$     & 66.67 $\pm$ 36.51 & 13 / 4 \\
PLA $15\%$     & 86.67 $\pm$ 26.67 & 13 / 4 \\
PLA $20\%$     & 86.67 $\pm$ 16.33 & 14 / 4 \\
\midrule
Beef           & 83.33 $\pm$ 21.08 & 12 / 4 \\
Chicken        & 80.00 $\pm$ 26.67 & 12 / 4 \\
Pork           & 90.00 $\pm$ 20.00 & 11 / 3 \\
\hline
\hline
\end{tabular}
\end{table}

\subsection{Hard Material Classification}\label{sec:material_cls_hard}
The first task undertaken was to identify very subtle changes in the internal infill density of common 3D-printed material. Circular plates of PLA with infill densities of $10\%$, $15\%$, $20\%$ were made for testing. Using the method described in \cref{sec:data_collection}, a 3x3 grid pattern was executed with $5$~mm spacing between each point. The target contact force was predetermined to be $3$~N, with a maximum allowable error of $1.5$~N. For each infill density, two trials were conducted on two different samples. As seen in  \cref{tab:cv_accuracy}, the classification accuracy has high variance for PLA $10\%$ potentially due to similar acoustic features as with other classes. %
This is further corroborated by \cref{fig:classification}a, where there is an overlap of frequencies for PLA $10\%$ and $15\%$ materials. The observation translates to a well-held-out test set accuracy of $75\%$ for PLA $10\%$ and $100 \%$
for PLA $15\%$ and PLA $20\%$. 

\subsection{Soft Material Classification}\label{sec:material_cls_soft}
\begin{figure}[t]
    \centering
    \includegraphics[width=0.98\linewidth,trim=2pt 2pt 2pt 2pt,clip]{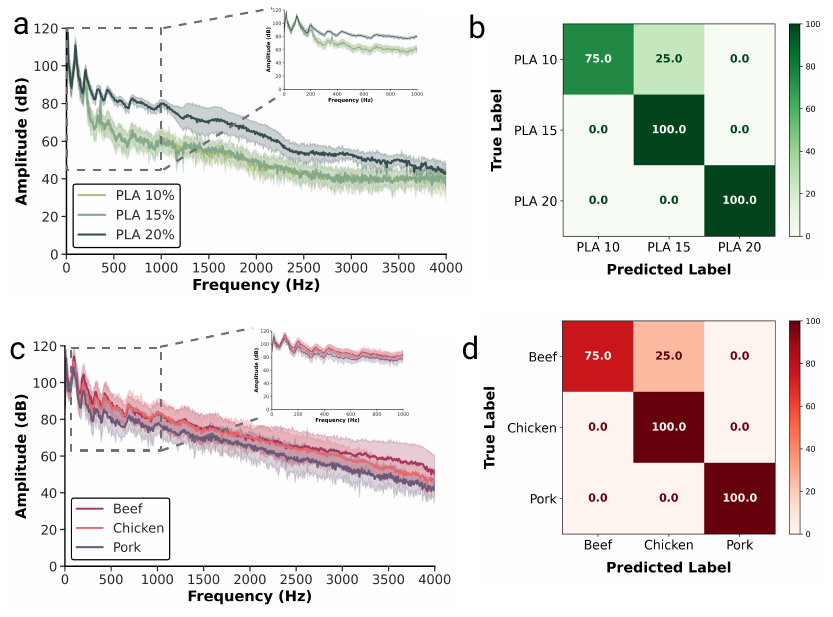}
    \vspace{-2mm} 
    \caption{\textbf{Material Classification}: (a) PLA material palpation frequency distribution with emphasis on 0--1000 Hz range; (b) Hard material classification result; (c) Soft-tissue material palpation frequency distribution with emphasis on 0--1000 Hz range; (d) Soft material classification result.}
    \label{fig:classification}
\end{figure}

For efficacy in surgical scenarios, excised \textit{ex vivo} tissues of bovine (beef), chicken breast, and porcine (pork) were used. A 4x4 grid of palpation patterns was used to maximize data point collection from each sample, with a spacing of $10$~mm between each point for maximizing variability in properties over the tissue sample. The target contact force was lowered to $2$~N to avoid puncturing or damaging the samples owing to softer materials. In \cref{fig:classification}c, it can be seen that beef and chicken tissue had similar lower frequency responses with higher frequency components diverging, while pork tissue's response was different. As shown in \cref{tab:cv_accuracy}, animal tissues perform much better than hard materials (\cref{sec:material_cls_hard}) as seen through 5-fold cross-validation accuracy. This is also reflected in \cref{fig:classification}d, with pork tissue achieving $100\%$ accuracy alongside chicken and beef having $75~\%$ accuracy.



\section{Discussion and Conclusion}\label{sec:discussion}

In this work, we propose PalpAid, a low-cost, modular tactile sensor for conformal tissue identification in RAMIS and general robotic applications. PalpAid consists of three main components: a flexible, expandable palpator, a high-SNR microphone, and a sensitive pressure sensor to gently probe tissues. The applied contact force deforms the palpator, generating internal waves in the confined air cavity that result in measurable differences in acoustic and pressure signals. The biocompatible design enables use in surgical applications as it is easy to replace the palpator and top cap after use between surgeries. This gives a unique advantage over several VBTS devices, which are costly to replace.

The sensor functionality relies on the volume of enclosed air in the palpator. Hence, the total time the valve remains open controls internal pressure, i.e, dimension and sensor performance. During experiments, it was seen that though the internal pressure gradually decreased over cycles of palpation, the trend remained consistent among trials and palpator sample versions, demonstrating consistency over manufacturing.
%

In theory, solely relying on an individual acoustic sensor or an individual pressure sensor for tissue classification should be possible, but it is challenging in practice. Each individual sensor (pressure and microphone) can only provide partial information about the contact event. 
Thus, this work aimed to bring these two modalities together by using pressure to indicate contact and audio signal during contact to classify material. 
This strategy mitigates the effects of environmental noise on the audio signal. The combined performance is demonstrated in \cref{fig:Tap_peak}b, where, though the spectrogram shows stronger signals in lower frequency bands between the 12th and 13th detected peaks, the absence of a corresponding pressure signal correctly marks the event as an outlier. Furthermore, PalpAid shows strong performance over both very hard (and stiff) and soft materials alike. Even in the presence of heterogeneous tissue layers and varied contours, the sensor is able to pick up minute changes in tissue properties for classification tasks. It is also worth noting that MFCC features are high-level and do not capture all information present in the acoustic signals, and provides validation for future work using learning-based audio encoders. 

While the device possesses benefits, there are several areas to improve upon for robust soft-tissue assessment in long-term-horizon tasks and increasing testing sample size. Firstly, the lack of closed-loop pressure feedback reduces control over internal pressure during inflation-deflation cycles; in the future, the control loop proposed by \cite{10521918} can be used to monitor and adjust the change in pressure over time. 
Secondly, in its current form, the large size of the palpator ($5$~mm diameter) reduces the resolution of tissue boundary identification and does not provide the sub-millimeter resolution standard for surgical applications. Further miniaturization could involve micromolding and more precise SLA manufacturing techniques, which would enable the sensor to be embedded within a surgical trocar. In current proof-of-concept form, the true multimodality of the sensor is not leveraged, as both pressure and acoustic sensor signals perform different actions. Future work should aim to combine information from both sensors for classification. Finally, during clinical deployment all components except the palpator would be covered with clinical drapes, preserving biocompatibility. Additional work should still be done to replace the epoxy-based materials with medical industry standards. 


In conclusion, we present a novel pneumatic tactile sensor that leverages complementary sensing modalities to capture high-frequency acoustic signals from gentle tool-tissue interaction for soft-tissue delineation in real-time. The low cost, minimal fabrication, and smaller form factor allow easy integration in current robotic and handheld surgical scenarios. 
In the future, experiments will be conducted on soft-tissue brain phantoms \cite{prakash2023brain} and tissues of known mechanical properties to quantify the sensitivity of the sensor for further clinical use. Additionally, a hardware-software co-design will be undertaken to reduce the sensor form factor and utilize state-of-the-art signal processing techniques, such as spectrogram representation with convolutional neural networks, to leverage the rich signals present in acoustic data fully. Future work would aim to identify common feature sets between pressure and acoustic signals, while exploring additional modalities such as vision, to inform tissue identification and boundary delineation. 



\section*{Acknowledgments}

The authors would like to acknowledge the help of Danyi Chen in setting up the palpation experiment and the members of the Brain Tool Lab for their feedback and comments.

\bibliographystyle{IEEEtran}  
\bibliography{IEEEfull}  

\begin{thebibliography}{10}
\providecommand{\url}[1]{#1}
\csname url@rmstyle\endcsname
\providecommand{\newblock}{\relax}
\providecommand{\bibinfo}[2]{#2}
\providecommand\BIBentrySTDinterwordspacing{\spaceskip=0pt\relax}
\providecommand\BIBentryALTinterwordstretchfactor{4}
\providecommand\BIBentryALTinterwordspacing{\spaceskip=\fontdimen2\font plus
\BIBentryALTinterwordstretchfactor\fontdimen3\font minus \fontdimen4\font\relax}
\providecommand\BIBforeignlanguage[2]{{%
\expandafter\ifx\csname l@#1\endcsname\relax
\typeout{** WARNING: IEEEtran.bst: No hyphenation pattern has been}%
\typeout{** loaded for the language `#1'. Using the pattern for}%
\typeout{** the default language instead.}%
\else
\language=\csname l@#1\endcsname
\fi
#2}}

\bibitem{gorman2021importance}
B.~G. Gorman, J.~Hanson, and N.~Y. Vidal, ``The importance of palpation in the skin cancer screening examination,'' \emph{Journal of Cosmetic Dermatology}, vol.~20, no.~12, pp. 3982--3985, 2021.

\bibitem{shetty2015accuracy}
D.~Shetty, B.~V. Jayade, S.~K. Joshi, and K.~Gopalkrishnan, ``Accuracy of palpation, ultrasonography, and computed tomography in the evaluation of metastatic cervical lymph nodes in head and neck cancer,'' \emph{Indian journal of dentistry}, vol.~6, no.~3, p. 121, 2015.

\bibitem{moschovas2024first}
M.~C. Moschovas, S.~Saikali, A.~Gamal, S.~Reddy, T.~Rogers, M.~C. Sighinolfi, B.~Rocco, and V.~Patel, ``First impressions of the new da vinci 5 robotic platform and experience in performing robot-assisted radical prostatectomy,'' \emph{European Urology Open Science}, vol.~69, pp. 1--4, 2024.

\bibitem{su2019multicamera}
Y.-H. Su, K.~Huang, and B.~Hannaford, ``Multicamera 3d reconstruction of dynamic surgical cavities: Non-rigid registration and point classification,'' in \emph{2019 IEEE/RSJ International Conference on Intelligent Robots and Systems (IROS)}.\hskip 1em plus 0.5em minus 0.4em\relax IEEE, 2019, pp. 7911--7918.

\bibitem{yuan2017gelsight}
W.~Yuan, S.~Dong, and E.~H. Adelson, ``Gelsight: High-resolution robot tactile sensors for estimating geometry and force,'' \emph{Sensors}, vol.~17, no.~12, p. 2762, 2017.

\bibitem{zhang2025utact}
Q.~Zhang, Z.~Zuo, H.~Wang, B.~Liu, Y.~Yilihamu, and L.~Wen, ``Utact: Underwater vision-based tactile sensor with geometry reconstruction and contact force estimation,'' \emph{Advanced Robotics Research}, p. 202500091, 2025.

\bibitem{tiong2025omnisense}
T.~J.~T. Tiong and A.~R. See, ``Omnisense v2: A human-skin inspired visuotactile sensor for unified tactile imaging,'' \emph{IEEE Sensors Journal}, 2025.

\bibitem{prince2025tacscope}
M.~R.~I. Prince, S.~Athar, P.~Zhou, and Y.~She, ``Tacscope: A miniaturized vision-based tactile sensor for surgical applications,'' \emph{Advanced Robotics Research}, p. e202500117, 2025.

\bibitem{di2024using}
J.~Di, Z.~Dugonjic, W.~Fu, T.~Wu, R.~Mercado, K.~Sawyer, V.~R. Most, G.~Kammerer, S.~Speidel, R.~E. Fan, \emph{et~al.}, ``Using fiber optic bundles to miniaturize vision-based tactile sensors,'' \emph{IEEE Transactions on Robotics}, 2024.

\bibitem{kapuria2024robot}
S.~Kapuria, J.~Bonyun, Y.~Kulkarni, N.~Ikoma, S.~Chinchali, and F.~Alambeigi, ``Robot-enabled machine learning-based diagnosis of gastric cancer polyps using partial surface tactile imaging,'' in \emph{2024 IEEE/RSJ International Conference on Intelligent Robots and Systems (IROS)}.\hskip 1em plus 0.5em minus 0.4em\relax IEEE, 2024, pp. 2360--2365.

\bibitem{liu2025sonicsense}
J.~Liu and B.~Chen, ``Sonicsense: Object perception from in-hand acoustic vibration,'' in \emph{Conference on Robot Learning}.\hskip 1em plus 0.5em minus 0.4em\relax PMLR, 2025, pp. 4332--4353.

\bibitem{chen2022boombox}
B.~Chen, M.~Chiquier, H.~Lipson, and C.~Vondrick, ``The boombox: Visual reconstruction from acoustic vibrations,'' in \emph{Conference on Robot Learning}.\hskip 1em plus 0.5em minus 0.4em\relax PMLR, 2022, pp. 1067--1077.

\bibitem{wall2023passive}
V.~Wall, G.~Z{\"o}ller, and O.~Brock, ``Passive and active acoustic sensing for soft pneumatic actuators,'' \emph{The International Journal of Robotics Research}, vol.~42, no.~3, pp. 108--122, 2023.

\bibitem{li2025acoustac}
M.~S. Li and H.~S. Stuart, ``Acoustac: Tactile sensing with acoustic resonance for electronics-free soft skin,'' \emph{Soft Robotics}, vol.~12, no.~1, pp. 109--123, 2025.

\bibitem{liu2025wildfusion}
Y.~Liu and B.~Chen, ``Wildfusion: Multimodal implicit 3d reconstructions in the wild,'' in \emph{2025 IEEE International Conference on Robotics and Automation (ICRA)}.\hskip 1em plus 0.5em minus 0.4em\relax IEEE, 2025, pp. 8603--8603.

\bibitem{mandil2024acoustic}
W.~Mandil, K.~Nazari, S.~Parsons, A.~Ghalamzan, \emph{et~al.}, ``Acoustic soft tactile skin (ast skin),'' in \emph{2024 IEEE International Conference on Robotics and Automation (ICRA)}.\hskip 1em plus 0.5em minus 0.4em\relax IEEE, 2024, pp. 4105--4111.

\bibitem{andrussow2025adding}
I.~Andrussow, J.~Solano, B.~A. Richardson, G.~Martius, and K.~J. Kuchenbecker, ``Adding internal audio sensing to internal vision enables human-like in-hand fabric recognition with soft robotic fingertips,'' in \emph{2025 IEEE-RAS 24th International Conference on Humanoid Robots (Humanoids)}.\hskip 1em plus 0.5em minus 0.4em\relax IEEE, 2025, pp. 01--08.

\bibitem{bao2025miniaturized}
E.~Bao, C.~Fang, and D.~Song, ``A miniaturized and low-cost fingertip optoacoustic pretouch sensor for near-distance ranging and material/structure classification,'' \emph{IEEE Sensors Journal}, 2025.

\bibitem{chen2025boundary}
Z.~Chen, A.~C. Cahilig, S.~Dias, P.~Kolar, R.~Prakash, and P.~J. Codd, ``Where is the boundary? multimodal sensor fusion test bench for tissue boundary delineation,'' in \emph{IEEE-EMBS International Conference on Body Sensor Networks 2025}, 2025.

\bibitem{prakash2025biorender}
R.~Prakash, ``Created in biorender,'' \url{https://BioRender.com/m0xbfmk}, 2025, created with BioRender.com.

\bibitem{janardhana2025comprehensive}
R.~Janardhana, F.~Akram, Z.~Guler, A.~Adaval, and N.~Jackson, ``A comprehensive experimental, simulation, and characterization mechanical analysis of ecoflex and its formulation under uniaxial testing,'' \emph{Materials}, vol.~18, no.~13, p. 3037, 2025.

\bibitem{SmoothOn_Ecoflex_00_20}
I.~Smooth-On, ``Ecoflex™ 00-20 product information,'' \url{https://www.smooth-on.com/products/ecoflex-00-20/}, 2025, accessed: 2025-10-30.

\bibitem{ai2012classification}
O.~C. Ai, M.~Hariharan, S.~Yaacob, and L.~S. Chee, ``Classification of speech dysfluencies with mfcc and lpcc features,'' \emph{Expert Systems with Applications}, vol.~39, no.~2, pp. 2157--2165, 2012.

\bibitem{10521918}
H.~C. Cheung, C.-W. Chang, B.~Jiang, C.-Y. Wen, and H.~K. Chu, ``A modular pneumatic soft gripper design for aerial grasping and landing,'' in \emph{2024 IEEE 7th International Conference on Soft Robotics (RoboSoft)}, 2024, pp. 82--88.

\bibitem{prakash2023brain}
R.~Prakash, K.~K. Yamamoto, S.~R. Oca, W.~Ross, and P.~J. Codd, ``Brain-mimicking phantom for photoablation and visualization,'' in \emph{2023 International Symposium on Medical Robotics (ISMR)}.\hskip 1em plus 0.5em minus 0.4em\relax IEEE, 2023, pp. 1--7.

\end{thebibliography}

\end{document}